\documentstyle[fleqn,prl,aps,rotate,amssymb,preprint,epsf]{revtex}

\begin{document}
\tightenlines
\draft
%\twocolumn
\widetext

\title{Self-similarity under inflation and level statistics: a study
 in two dimensions}
\author{A. Jagannathan}
\address{
Laboratoire de Physique des Solides,
Universit\'{e} Paris--Sud,
91405 Orsay,
France }
\maketitle

\begin{abstract}
Energy level spacing statistics are discussed for 
the octagonal tiling, a two dimensional quasiperiodic structure.
A recursion relation is written for the probability
distributions of variables
defined on finite size approximants to this quasiperiodic tiling,
using their property of similarity under inflation.
Three types of distribution
functions are introduced and determined by a combination of numerical
and analytical techniques - these are likely to be of general 
utility in systems lacking translational invariance but
with inflation symmetry.
\end{abstract}

\narrowtext
Electrons in quasi-periodic solids are expected to have electronic properties
that are in some sense intermediate between those of perfect crystals and
disordered solids. This comes from
noting that on the one hand, the lack of periodicity of the tilings precludes
construction of extended Bloch states as in the case of crystals, thus
reducing the possibility of metallic behavior or these systems. On the other
hand, in a strongly disordered solid, electronic states existing for a
given energy in a given neighborhood of the sample are located exponentially
far away from a 
similar solution, 
whereas in the quasiperiodic 
tiling, similar environments are guaranteed to occur within a much
shorter distance -- about twice the linear size of the region --
 thus allowing a greater overlap.
The quasicrystal wavefunctions
are probably "critical" states, with a power law long distance behavior
and truly localized states should exist only exceptionally 
 (such as the ring-
states for special energies mentioned further below).  
Although one dimensional quasiperiodic systems have been successfully
tackled by renormalization group methods \cite{koh}, 
Hamiltonians in two and higher dimensions
have been mainly examined numerically, for example
in the 2D Penrose and octagonal tilings \cite{ben}.

It is known in the case of disordered Hamiltonians 
that level spacing statistics and the underlying wavefunctions 
fall into two categories :
Wigner-Dyson gaussian statistics associated with extended electronic 
wavefunctions in the case of weak disorder, and Poissonian statistics
associated with exponentially localized wavefunctions in the case of strong 
disorder. Although quasicrystals are not necessarily disordered,
the complexity of the Hamiltonian makes it necessary and
useful to carry out such statistical analyses, perhaps to discover
 a new type of statistical behavior in these media.

In the work now to be discussed, we have used the 
inflation symmetry \cite{lev}
of quasiperiodic systems to define a set of recursively related probability
distributions for level statistics.
Inflation refers to the (reversible) operation in which a subset
of vertices of the tiling are erased, and the new vertices reconnected
according to a well-defined prescription.
 The infinite quasicrystal is left unchanged
by this transformation, aside from the trivial multiplicative change of edge
lengths by a factor $\lambda$ and a corresponding
reduction in the vertex density 
 by $\lambda^d$ in $d$ dimensions. For the octagonal tiling, this
factor is $\lambda = 1 +\sqrt 2$. This symmetry was used in \cite{moss}
in a discussion of the spectrum of tight-binding models on the octagonal
tiling, a full renormalization
scheme (such as in \cite{pjb} for the
Fibonacci chain) has not yet been obtained for the 2D tilings.
 Previous numerical studies did not take
this property of inflation into consideration in analysing level
statistics. However inflation symmetry is an important symmetry 
in the absence of translational invariance  and
the recursive scheme proposed in this paper is an attempt to utilise
this property in solving for the level spacing probability distribution. 

The systems that we consider are square pieces of tiling
whose interiors
reproduce the quasiperiodic arrangement found in the perfect infinite
octagonal tiling, and with edges that permit the periodic repetition
along the two directions of the plane. These "square approximants" of the
octagonal tiling can be obtained by projection from a four dimensional
cubic lattice by the "cut-and-project" method \cite{dunog}.
Fig.1a shows the $k=3$ approximant, containing 239 sites, while
 larger square
approximants are generated using the series of 
rational approximants: 
$\lambda_k^{-1} = \frac{1}{2}, \frac{2}{5}, \frac{5}{12}...$.
 Here we present results obtained for 
 $k=4,5,6$,
containing 1393, 8119 and 47321 sites respectively (the number of sites $N_k$
increases by approximately $\lambda^2$ with each increase in $k$).
Successive approximants are related to each other by
an inflation operation. The tight-binding Hamiltonian is
defined by taking a hopping matrix element $t=1$ between connected sites
(see Fig.1), and imposing periodic boundary conditions. The resulting sparse
matrices were diagonalized by Lanczos routines to obtain the energy 
levels. Because of the presence of a discrete symmetry (reflection about
the line x=y) it turns out that adding a Bohm-Aharonov flux does not
change the model from  the orthogonal to unitary symmetry class
\cite{comm}. This fact was used
to augment the statistics for the smaller approximants by varying
the boundary conditions and obtaining a large number of spectra. 
 The level spacings are calculated independently for each subspace,
we present here the results for the symmetric subspace (even parity
under x- and y- translations and reflection across the diagonal), with
21750 nondegenerate levels for the biggest size considered.
Generally speaking, all the levels are in fact 
non-degenerate, aside from a macroscopic 
degeneracy of the $E=0$
states in this model, due to the eightfold ring-states that are permitted
in this geometry - Fig.1 illustrates such a wavefunction:
localized on the eight highlighted sites whose
wavefunctions are of constant magnitude but
alternating sign - this degeneracy is well-understood
and does not affect the level statistics.
%\begin{figure}[htb]
%\centerline{\epsfysize 12cm
%\rotate[l]{\epsffile{fig1.ps}}
%}
%\end{figure}

Fig.2) shows a typical density of states (DOS) histogram obtained for the
approximants.
Its characteristically spiky shape poses the problem
of "unfolding" the level spacings: in conventional random
systems, fluctuations of the density of states (DOS)
 remain small enough with respect to 
 a more smoothly varying underlying contribution, which is
then used in renormalizing or unfolding the level spacings. In the randomized
versions of this model that were considered in \cite{pj}
this unfolding $is$ possible as the density of states has an
underlying smooth part. But 
in the absence of randomness, the DOS has $no$ smooth variations, and 
the limit
$\rm lim_{N \rightarrow \infty} \rho (E)$ for fixed $E$ does not exist: the
"fluctuations" are of the same order as the "average" DOS and they are present
on arbitrarily small energy scales.
We consider first therefore the
``bare" (i.e. sans unfolding) level statistics for these systems. 
Fig.3a shows the numerically calculated points and
the fitted Lognormal (LN) distribution first reported in \cite{pj}
for the biggest approximant and Fig.3b shows
the points obtained for three different system sizes, after shifts
  so as to be centered at the origin. The
points lie on gaussians of the form
$\tilde{P}_k({\rm ln}s) = \sqrt{\frac{\pi}{\alpha}} {\rm e}^{-
\alpha ({\rm ln}s - a_k)^2}$. The peak position  $a_k$ shifts 
with increasing $k$, reflecting the fact that
the mean value of $s$ is reduced by $\lambda^2$.
More surprisingly, the width does not seem to depend very much on
system size. The fitted values $\alpha$ 
vary from about 0.79 to 0.81 from the smallest to largest system size,
with an uncertainty of about 0.01 due to the fluctuations and the
deviation from gaussian behavior in the tails.
The  continuous curve in Fig.3b is obtained from the
recursive calculation discussed further below.

 LN distributions may be obtained from iterative
 schemes such as the following: at
each step of the iteration
 intervals are split into two nonequivalent
subintervals with some choice of renormalisation
factors.  After $n$ steps, for $n$ sufficiently large, one approaches a
LN distribution of interval lengths as the saddlepoint
approximation used becomes better (such a calculation
was done in \cite{arc} for a resistive network).
This heirarchical construction is appealing
 for quasiperiodic systems, with their inflation-invariant
 structure and would predict that both
  the mean value as well as the variance
of ${\rm ln}s$ are linear in $n$ .
However since this latter behavior is $not$ observed ,
 the distribution of spacings for the quasicrystal
must be more compact than this simple model predicts, implying a more
``rigid" spectrum.

We now examine unfolding procedures to correct the level spacings
for DOS variations. A first attempt was reported in \cite{these} where each
spacing was divided by a local mean value of the level spacing. One can set
 $s_i^{(m)} = (E_{i+1} - E_{i})/(\frac{E_{i+m} - E_{i-m}}{2m})$
where the ``most local" unfolding corresponds to taking $m=1$.
The resulting variables $\{s^{(m=1)}\}$ were found to
follow a Wigner-Dyson distribution. As $m$ is increased, the 
probability distribution of the variables $s^{m}_i$
 shifts continuously over to
the LN distribution.
This result, that
the level spacing obeys W-D statistics upon unfolding was rediscovered
by Zhong et al \cite{zh} where they smoothed the integrated DOS using
a spline fit. This works because the IDOS, while not a smooth curve, has
eigenvalues that are grouped together in a way as to allow a piecewise
continuous fit -
 this procedure yields the same result as in \cite{these} for m
small. 

To resume, 
the spacing statistics are W-D when the
unfolding procedure is local, i.e. spacings
are measured by the yardstick of the local DOS. At the
other extreme, when the
unfolding  is done using the global mean level spacing 
the LN distribution is
obtained. These unfolding schemes do not make use of the inflation
transformation relating the spectra of tilings of successive sizes. 
With each inflation, there are $\lambda^2$ times more levels
  $E^{(k)}_{i}$, which fall within the
energy intervals of the smaller system in an inhomogeneous fashion. 
  Otherwise put, each  "ancestor" interval $s_i$ 
of the $(k-1)$th approximant
splits into a certain number $n_i$ of intervals of 
the (larger) $k$th approximant,
with the mean value of $n_i$ being equal to $\lambda^2$.
We propose thus to renormalise each level spacing of the larger system
by a mean local spacing, 
 defined with respect to the common ancestor as follows:
\begin{equation} \label{eq:rec}
x^{(k)}_i = \frac{s^{(k)}_i} {\langle s \rangle _j} \equiv
      \frac{E^{(k)}_{i+1} - E^{(k)}_i}{(E^{(k-1)}_{j+1} - E^{(k-1)}_j)/
      n_j}
\end{equation}
where the indices $i$ and $j$
run from 1 to $N_k-1$ and $N_{k-1}-1$.  
$i$ indexes a given spacing of the bigger system, while
 $j$ is the index of the ancestor spacing which has split into
$n_j$ subintervals after inflation.  
to the $(k-1)$th approximant. Fig.4 shows the probability distribution
of the new variables $x^{(k)}_i$
for the case $k=5$ along with the Wigner-Dyson curve
$P_{WD}(x)=\frac{\pi}{2}x {\rm e}^{-\pi x^2/4}$.
 We note that the $x$ variables as defined 
automatically have the mean value of unity, and that {\it no
parameters} are involved in the fit to W-D statistics. For the square
approximants, one can thus define variables that correspond to the
unfolded level spacings and which have Wigner-Dyson statistics independently
of the system size (for systems large enough so as to provide
 reasonable statistics). The fit to $P_{WD}$ is not as good as
in \cite{these,zh}, due to the fact that the
transformation $k \rightarrow k+1$ is a discrete transformation -
if it were possible to tune $\lambda$ gradually up from unity to its
final value, one would have an unfolding that was truly local at each
step and the resulting $x-$distributions would 
be exactly $P_{WD}$.

Consider Eq.\ref{eq:rec} after one more iteration:
\begin{equation} \label{eq:rec2}
x_i^{(k)}x_j^{(k-1)} = n_j n_l \frac{ s_i^{(k)}}{s_l^{(k-2)}} 
\end{equation}
where $s^{(k)}_i$ is a subinterval of $s^{(k-1)}_j$ (subdivision index
$n_j$), which in turn is
a subinterval of $s^{(k-2)}_l$ (subdivision index $n_l$). The
$x$ variables on the $k$th and the $(k-1)$th levels obey the
 WD statistics.

The quantities that contain information about this
particular quasiperiodic system are the subdivision factors
$n_j$.
The distribution of $n_j$ is not a smooth function as the variable
takes on only a very limited set of values. However, when the inflation
is carried out twice, the resulting subdivisions have a wider spread of
values and a new distribution can in fact be defined.
The distribution of the product of subdivision factors appearing in
Eq.\ref{eq:rec2}
 was calculated for our set of three tilings,
with the result shown in Fig.5 ($n$ is the product variable). 
We propose the functional 
 form $Q_{n}^{(2)} = \beta^{2} n 
 {\rm e}^{-\beta
n}$, for the analytical computation that follows,
where
$\beta^{-1} = \lambda^4 \approx 34$ (shown as a continuous
line in Fig.5). This implies that the
distribution of splitting factors follows
a Poisson distribution when $k \rightarrow k+2$  as if levels are
randomly distributed within the bandwidth. 
Although a theoretical calculation for this distribution
is lacking, one may speculate that
 it probably holds for other inflation
symmetric systems as well, with an appropriate choice of
the parameter $\beta$.
Note that the fluctuations around the smooth curve
in Fig.5 are still rather large
- whereas for higher products $n=\Pi_{i=1}^p n_i$ 
($p>2$) more continuous curves will be had, enabling 
 better fits of the corresponding 
distributions $Q_{n}^{(p)}$. 
 
According to Eq.\ref{eq:rec2} the level spacings at the $k$th and 
$(k-2)$th level 
approximants are related by
${\rm ln}s^{(k)} - {\rm ln}s^{(k-2)} = {\rm ln}(\frac{x^{(k)}
x^{(k-1)}}{n})  \equiv  Y$.
The distribution function of $Y$, $P_Y(Y)$, 
is thus given by the convolution of
the distribution functions of ${\rm ln}s^{(k)}$ and ${\rm ln}s^{(k-2)}$:
in terms of the Fourier transforms, $P_Y(q)=
\tilde{P}^{(k)}(q) \tilde{P}^{(k-2)}(-q)$. We find the distribution
of the product $x^{(k)} x^{(k-1)}/n = X = {\rm e}^Y$ by first
integrating over the $x$ variables, finding
\begin{equation} \label{eq:k0}
P_X = cst \quad \int {\rm d}n \quad n^2\quad Q_{n}^{(2)}
\quad K_0(\pi X/2)
\end{equation}
where $K_0$ is the zeroth order Bessel function. Putting in the
explicit form of $Q_{n}^{(2)}$, one obtains
\begin{equation} \label{eq:ln}
P_Y = cst \quad \frac{\tilde{X}^2}{(1 +  \tilde{X})^4}
\quad F(4,\quad\frac{1}{2};
\quad\frac{9}{2};\quad
\frac{1  - \tilde{X}}{1 +  \tilde{X}}) 
\end{equation}
where $F$ is the hypergeometric
function ($\tilde{X} = \pi X/(2\beta)$).
This function is very well approximated by
a gaussian form for $P_Y(Y)$, for $Y$ close to its
 most probable value 
 $Y_{max} \approx -4.1 $. One finds upon expanding around $Y_{max}$ that 
\begin{equation} \label{eq:ln2}
P_Y(Y) = cst \quad {\rm e}^{-\sigma (Y-Y_{max})^2}
\end{equation}
where $\sigma = 0.42$. This
 along with the convolution relation $P_Y = \tilde{P}^{(k-2)}
\tilde{P}^{(k)}$
 leads us to propose a gaussian form for the level spacings, 
\begin{equation} \label{eq:ln3}
\tilde{P}^{(k)}({\rm ln}s) =
 cst\quad {\rm e}^{-\alpha_k ({\rm ln}s - \langle {\rm ln}s
\rangle_k)^2} 
\end{equation}
with a similar relation with $k$ replaced by $k-2$ for $\tilde{P}^{(k-2)
}$. From the convolution relation for $P_Y$ one has
\begin{eqnarray} \label{eq:ln3}
P_Y(Y) = cst\quad  {\rm e}^{-\frac{1}{4} (\alpha_k +\alpha_{k-2}) (Y - 
\langle{Y}\rangle)^2}
\end{eqnarray}
where $\langle{Y}\rangle = \langle{\rm ln}s\rangle_k - 
\langle{\rm ln}s\rangle_{k-2}$. Comparing Eqs.\ref{eq:ln2} and \ref{eq:ln3}
for $P_Y$, one gets
\begin{eqnarray}
\frac{1}{4} (\alpha_k + \alpha_{k-2}) = \sigma  \quad ; \quad
\langle {\rm ln}s \rangle_{k-2} - \langle {\rm ln}s \rangle_k = Y_{max}
\end{eqnarray}
Thus if $\alpha$ is independent of $k$, one has $\alpha = 0.84$, the value
taken for the gaussian curve plotted in Fig.3b. 
The numerical data shown in Figs.3 are in accord with this, if one notes
that the fitted gaussian width increases slightly with system size
and approaches the value given above. The shifts calculated numerically
 were $\langle {\rm ln}s\rangle_{k+1} - \langle {\rm ln}s\rangle_k = 1.8$
for $k=4$ and $k=5$, while the expected value from the calculation
above is $\frac{1}{2} {\rm ln}Y_{max} \approx 2$.

The LN distribution is clearly only valid for values of ${\rm ln}s$
close enough to the mean value: the deviations from the gaussian
approximation to the exact expression in Eq.\ref{eq:ln} are most evident 
in the small ${\rm ln}s$ tail of Fig.3a.

In conclusion,  
level statistics in our quasiperiodic system are described by
both Wigner-Dyson statistics and the LN statistics depending on the
choice of unfolding used. The former indicates
that the model belongs in the class of nonintegrable models
 described by random matrix theory, while
 the latter has its origin in the inflation-symmetry of the tiling.
It would be interesting to
consider related models
 such as the Penrose tiling (with inflation parameter $\tau = 
\frac{\sqrt{5}-1}{2})$, where similar results should hold with
numerical differences only in the width
and the peak shifts of the LN distribution of spacings.

{\bf Acknowledgments}
I would like to thank IDRIS for computing time for this work.
I thank B. Altshuler and I. Aleiner for useful discussions in which
recursive schemes for probability distributions on successive approximants
were first considered.

%\bibliography{tiling}
%\bibliographystyle{prsty}

\begin{figure}[htb]
\centerline{\epsfysize 12cm
\rotate[l]{\epsffile{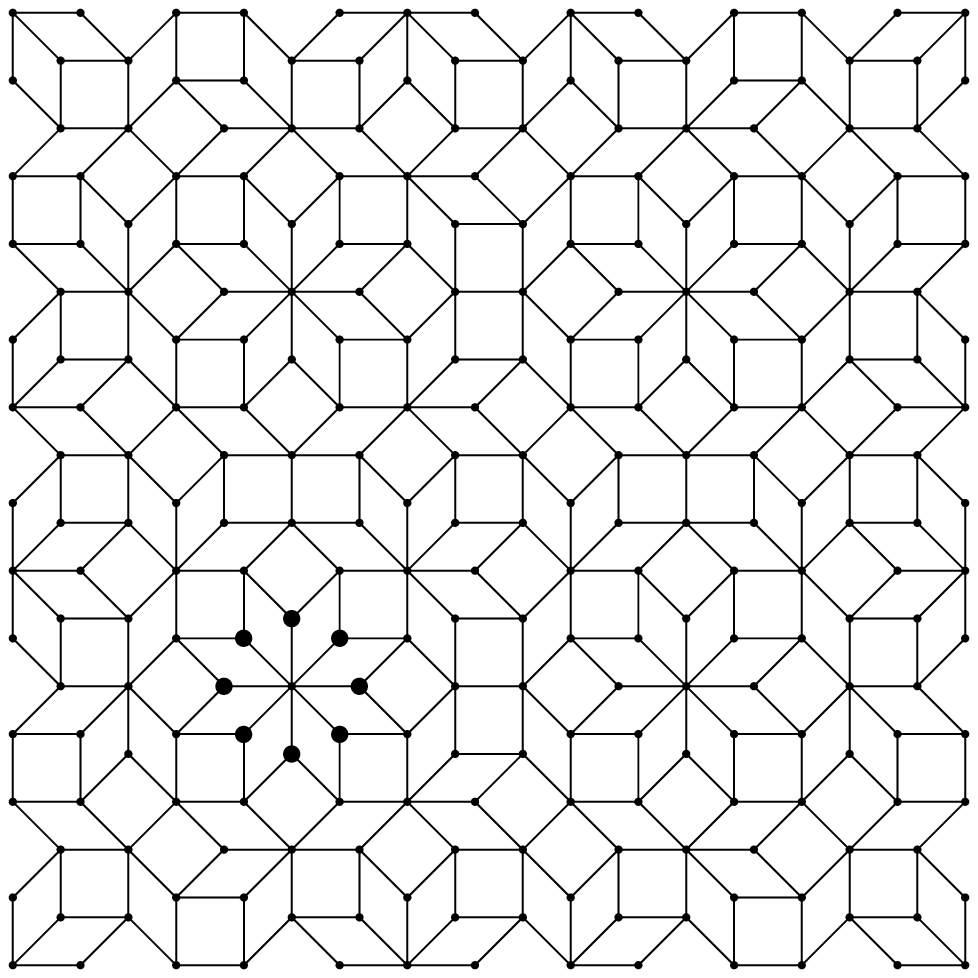}}
}
\caption[a]{Square approximant containing 239 sites, showing the eight-
fold symmetric sites around which the $E=0$ wavefunctions are
localised (see text).}
\label{f1}
\end{figure}

\begin{figure}[htb]
\centerline{\epsfysize 12cm
\rotate[l]{\epsffile{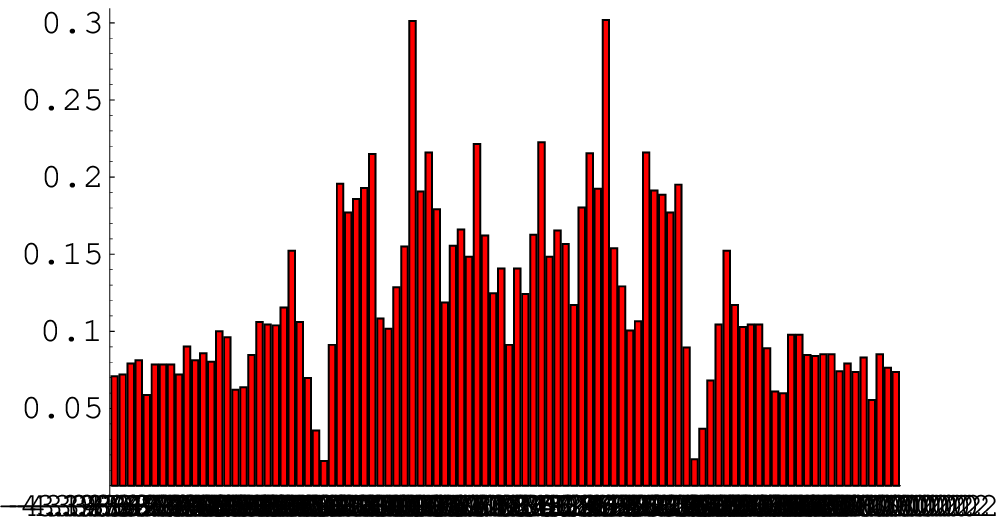}}
}
\caption[b]{Histogram of the density of states of a square approximant}
\label{f2}

\end{figure}

\begin{figure}[htb]
\centerline{\epsfysize 12cm
\rotate[l]{\epsffile{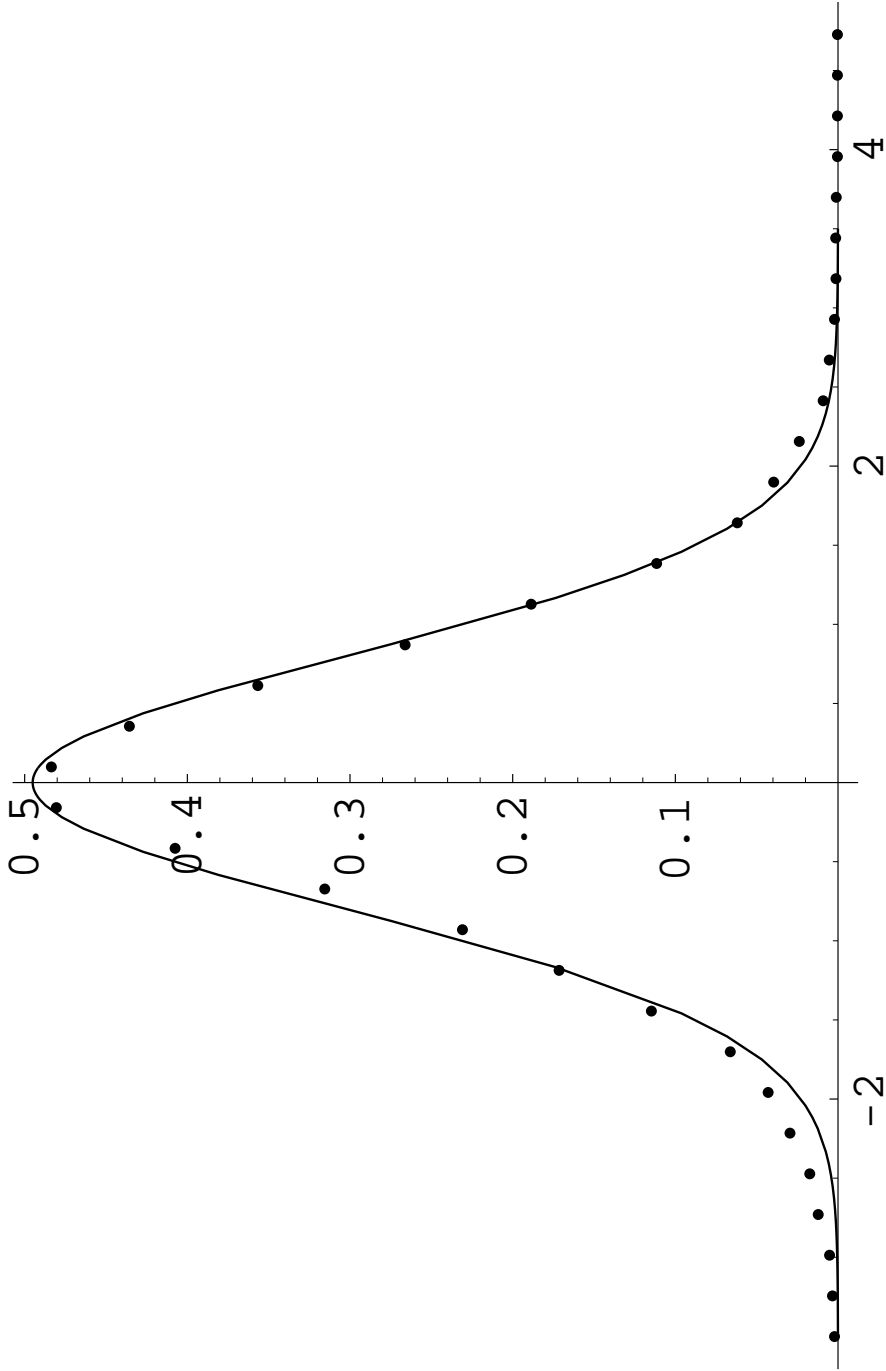}}
}
\caption[c]{Plot of $\tilde{P}({\rm ln}s)$ for the largest (k=6) approximant.
The continuous line is a gaussian fit.}
\label{f3}
\end{figure}

\begin{figure}[htb]
\centerline{\epsfysize 12cm
\rotate[l]{\epsffile{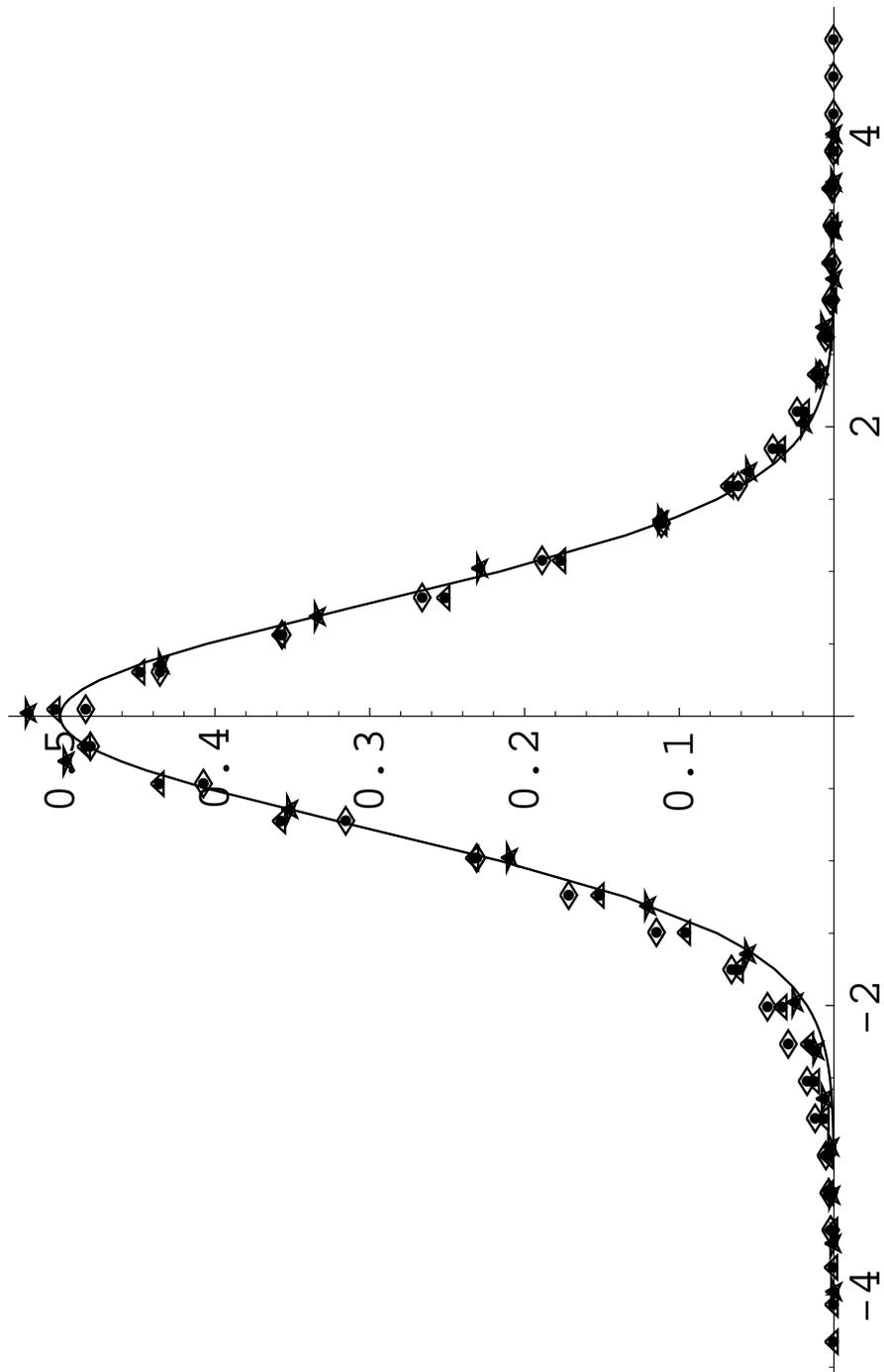}}
}
\caption[d]{Comparison of $\tilde{P}$ obtained for three approximants (k=4,5,6)
and theoretical curve (see text).}
\label{f4}

\end{figure}

\begin{figure}[htb]
\centerline{\epsfysize 12cm
\rotate[l]{\epsffile{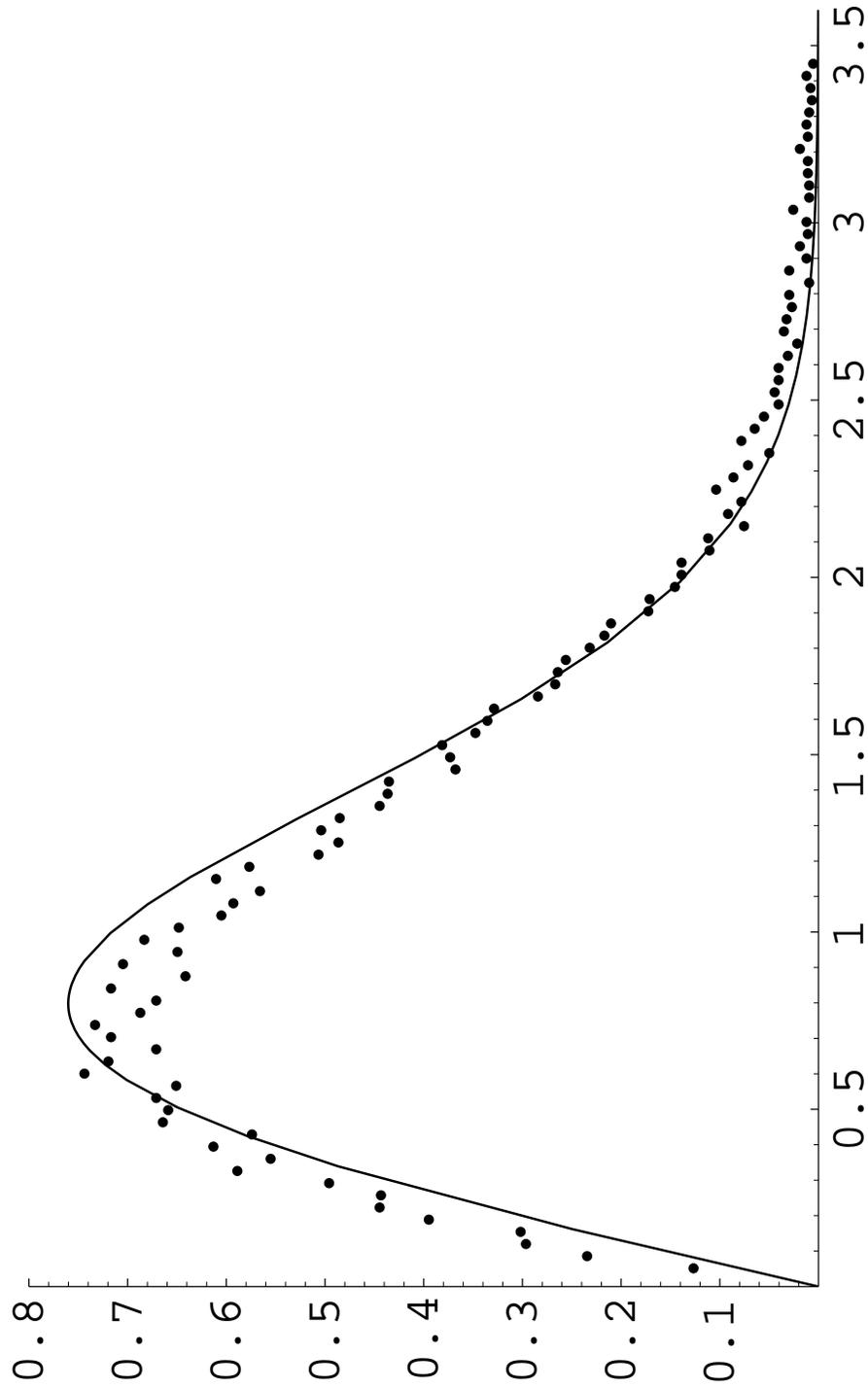}}
}
\caption[e]{Probability distribution of the unfolded level spacings using
Eq.\ref{eq:rec}. The points are numerical data, the curve is the
Wigner-Dyson law $P_{WD}$.}
\label{f1}
\end{figure}

\begin{figure}[htb]
\centerline{\epsfysize 12cm
\rotate[l]{\epsffile{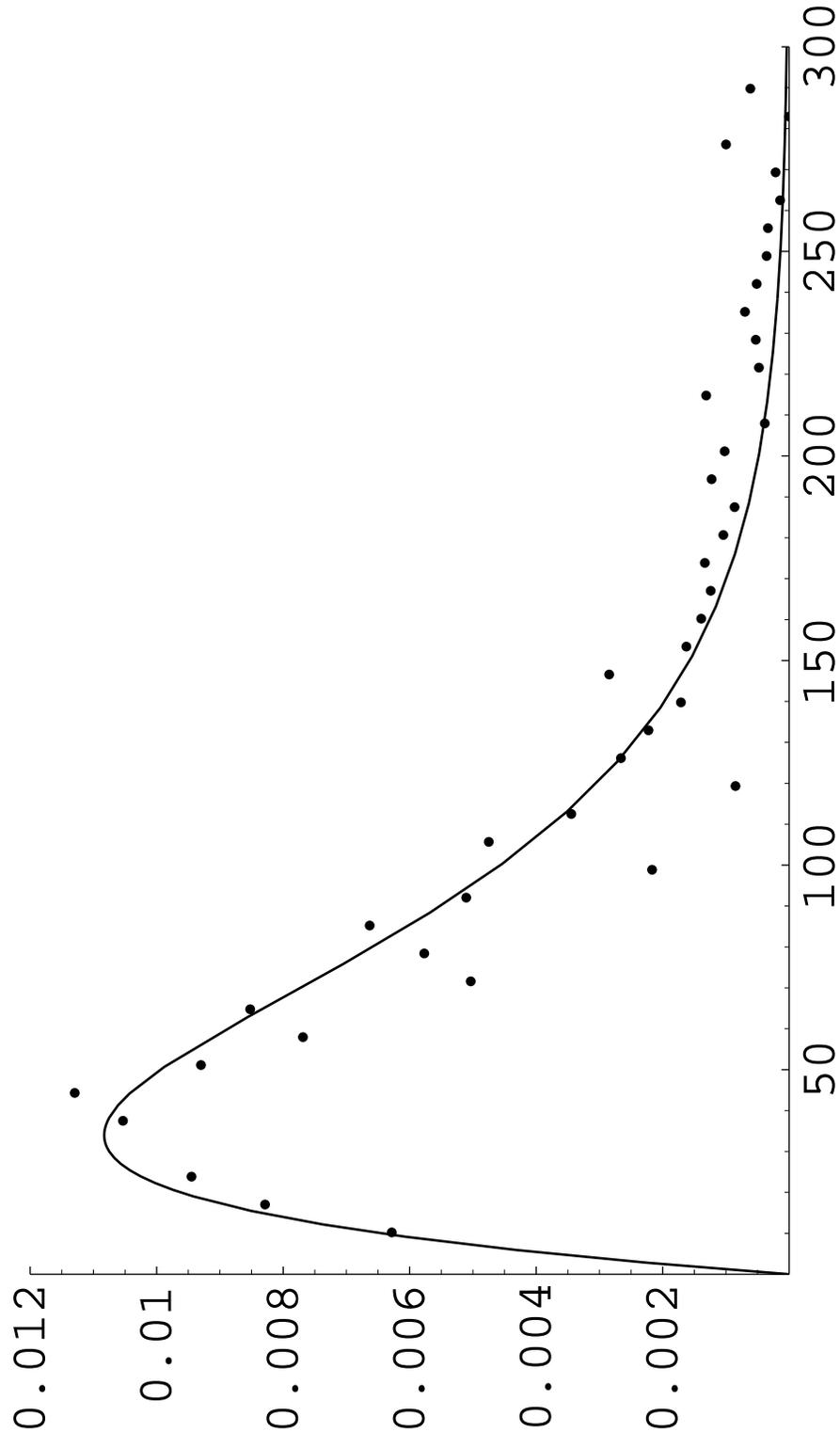}}
}
\caption[f]{Plot of the distribution of values of the product $n_in_j$ of
subdivision indices in going from the $(k-2)$th to $k$th approximant.
The curve is the analytical form discussed in the text.}
\label{f1}
\end{figure}

\end{document}